\documentstyle[twoside,fleqn,art10]{actaps}
\def\autor{R. Lietava}
\def\shorttitle{Simple model}
\begin{document}
\title{{A SIMPLE  MODEL FOR TRANSVERSE ENERGY DISTRIBUTION IN 
HEAVY ION COLLISIONS}}
{\bf Dedicated to J\'an Pi\v s\'ut on the occasion of his 60th birthday}
\author{R. Lietava}{Dept. of Theoretical Physics, 
Comenius University, Bratislava, Slovakia, \ \ \ \
CERN, Geneva, Switzerland}

\abstract{
A simple geometrical model (often quoted in literature as the Glauber model) 
of heavy ion collisions
is recapitulated. It is shown that the transverse
energy distribution of heavy ion collisions follow the geometry of 
the collision.
An extension of the model to include rapidity and transverse mass
particle spectra is discussed.}

\section{Introduction}  
Detailed studies of nucleus-nucleus interactions require the use of 
detailed event generators such as QGSM, RQMD or VENUS.  These models 
incorporate many features, and it can be difficult to understand the 
origin of relations between variables.  In the following, we describe
a simple geometrical model relating the number of collisions to the
cross section.  

We will assume a Glauber model. This model is widely used
in heavy ion community 
\cite{bialas}-\cite{rl}. 
Present
estimates of the collision centrality by the WA97 and NA50 collaborations
are based on this kind of model.
The physical ideas behind the model are  found
in \cite{the}.

\vspace {10pt}
\section{Model}    

\vspace{6pt}

A good introduction to the model is given in \cite{wong}.
Here we recapitulate the basic assumptions of the model.
The nucleus-nucleus collision is described in terms of nucleon-nucleon
collisions.
The nucleons of the nucleus with mass number $A$ are distributed in
nucleus with probability density $\rho(\vec b,z)$ normalised to unity.
It is convenient to introduce the thickness function for a
nucleus $A$ as
$$
T_A(\vec b)=\int \rho(\vec b,z)dz
$$
If the nucleus $A$ collide with a nucleus $B$ at impact parameter
$b$ the probability of having at least one 
inelastic nucleon-nucleon collision is
$$
p=\sigma_{in}\int T_A(\vec s)T_B(\vec b - \vec s) d^2\vec s
$$

The probability for the occurrence of $n$ inelastic baryon-baryon collisions
in the A-B collision is then
$$
P(n,\vec b) = \left( \begin{array}{c}AB\\
 n \end{array}\right) p^n(1-p)^{AB-n}
$$

It is easy to represent above formulas by a Monte-Carlo
program:

\begin{itemize}
\item the nucleons in nucleus $A$ are generated according
the probability density $\rho_A(\vec b,z)$
\item the nucleons in nucleus $B$ are generated according
the probability density $\rho_B(\vec b,z)$
\item nucleons from nucleus $A$ are ``collided'' with nucleons 
from nucleus $B$. The collision of nucleon $i$ in nucleus $A$ at
coordinates $(\vec b_A^i,z_A^i)$ with nucleon of nucleus $B$ at
coordinates $(\vec b_B^j,z_B^j)$ takes place if
$\vert \vec b_A^i-\vec b_B^j \vert < \sqrt{(\sigma_{in}/\pi)}
$
\end{itemize}

The number of participants $N_{part}$ (sometimes they are called 
``wounded nucleons'') is defined as the number of
nucleons in the target and projectile which have at least
one inelastic collision. 
The number of participants at impact parameter $\vec b$ is
$$
N_{part}=A\int d^2\vec s T_A(\vec s)\left( 1-(1-\sigma_{in}T_B(\vec s-\vec
b))^B\right) +\hfil
$$
$$
B\int d^2\vec s T_B(\vec s-\vec b)\left(1-(1-\sigma_{in}T_A(\vec s))^A\right).
$$

The number of collisions is the number of binary nucleon-nucleon collisions.
The number of collisions at impact parameter $\vec b$ is
$$
N_c=AB\sigma_{in}\int T_A(\vec s)T_B(\vec b - \vec s) d^2\vec s
$$

Both the number of participants and the number of collisions are
easily calculated using a Monte-Carlo code.

\vspace{6pt}
\section{Relation with Transverse Energy}
\vspace{10pt}
It is not possible to relate the geometrical model described above to
physically observable quantities without further assumptions.
To proceed we shall make two assumptions both based on empirical observations:
\begin{enumerate}
\item The number of produced particles is proportional to the number of participants
\item The number of produced particles is proportional to the transverse energy
in a given phase space window.
\end{enumerate}
The NA35
collaboration has studied multiplicity distributions in SS collisions
using detailed simulations in terms of the FRITIOF and VENUS models
\cite{bachler}.  In both models they find that the average charged 
multiplicity per participant is approximately constant as a function
of the number of participants.  
The simple proportionality of $<n_->$ and $<N_{p}>$ was also assumed by Bia{\l}as in
the wounded nucleon model \cite{bialas} and a similar 
result is obtained in the Dual Parton Model
\cite{DPM}. Comparison of several detailed models (FRITIOF, DTNUC 1.02,
VENUS 4.12) with data is done in \cite{ivan}.
The proportionality between the transverse energy and number of produced
particles was discussed in \cite{blasiuk}.

As an example of measured transverse energy distribution,
fig.1 shows the differential cross section of the transverse energy
produced in Pb-Pb and S-Au collisions at central rapidity as measured by 
NA49 collaboration \cite{na49}. 
\begin{figure}[t]
\begin{center}\mbox{\input epsf \epsfysize 7cm
                        \epsfbox{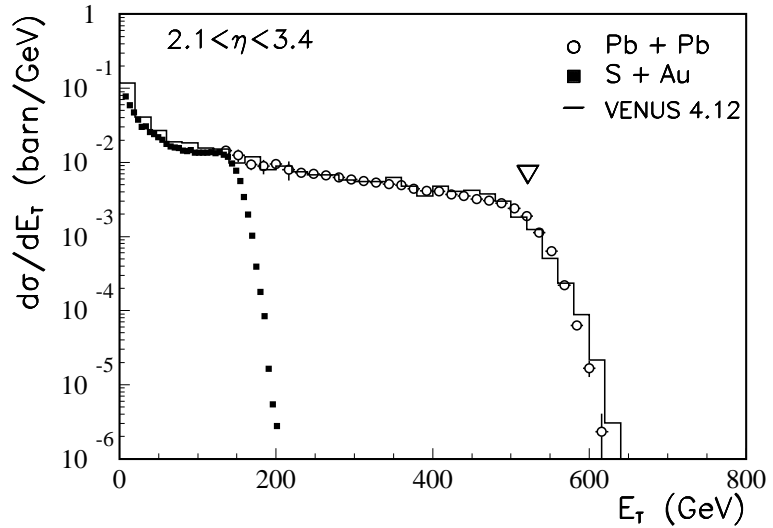}}\end{center}
\caption{Fig1 1. Differential cross section of the transverse energy
produced in Pb-Pb and S-Au at central rapidity 
as measured by NA49 coll. \cite{na49}}
\end{figure}   

The differential cross section of the number of participants for
the Pb-Pb and S-Au collisions calculated in our model is shown
in fig.2. The density of the nucleons in nucleus was taken to
be uniform in a sphere of radius $R=1.2A^{1/3}$. The inelastic 
cross section is $\sigma_{in}=30$ mbarn.
 Comparing fig.1 and fig.2 it is seen that energy per
participant is about 1.5 GeV.  This number fixes the 
y-scale absolutely.
Knowing that the number of generated events in fig.2 was $10^6$
and that they were generated in a circle with the area 6.35 barn
we see that the agreement of the cross section
(measured in barn/GeV) is good.

\begin{figure}[t]
\begin{center}\mbox{\input epsf \epsfysize 7cm
                        \epsfbox{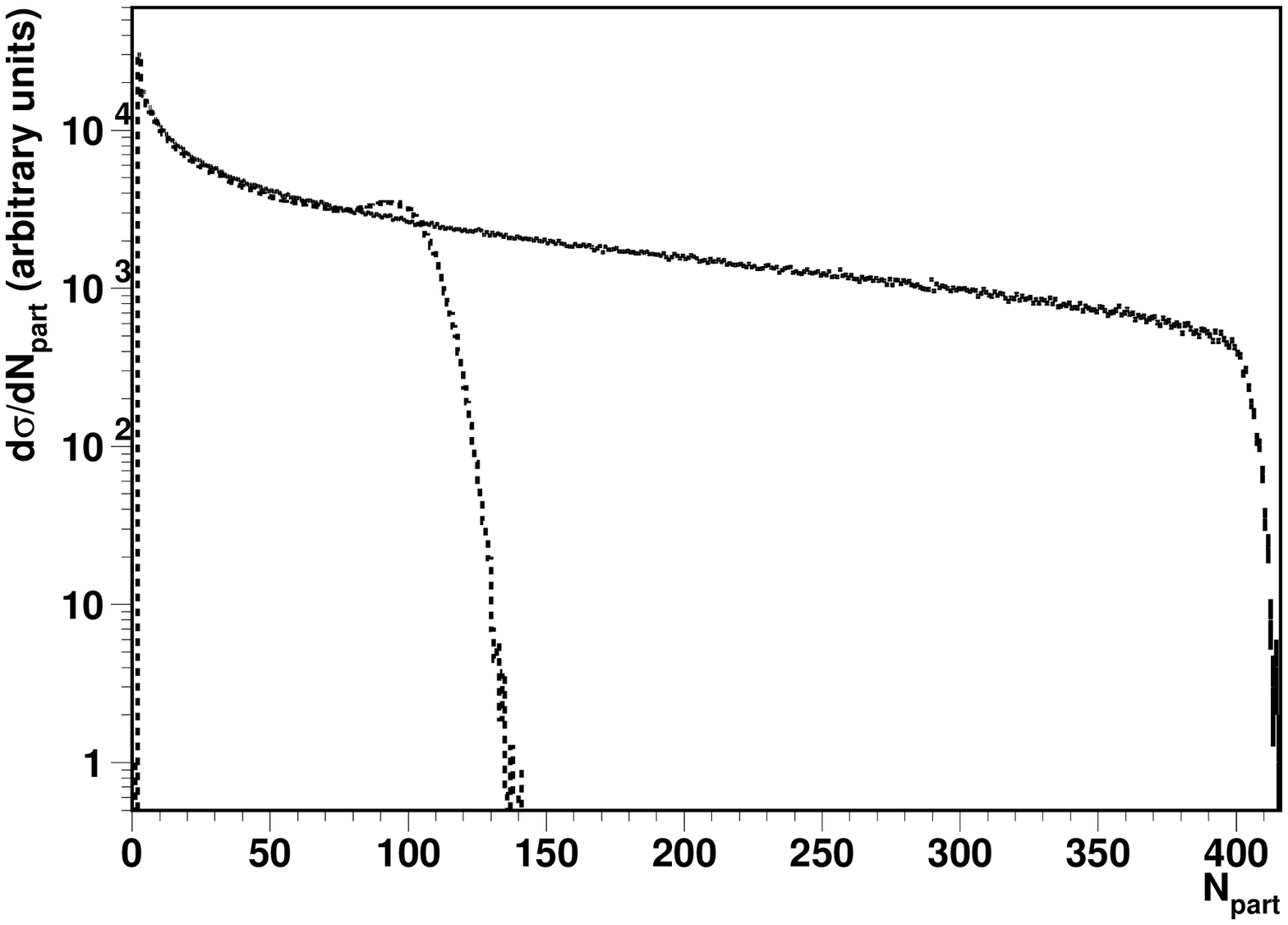}}\end{center}
\caption{Fig 2. Differential cross section of the number of 
participants in Pb-Pb collisions (crosses) and S-Au collisions
(histogram)}
\end{figure} 

We also present a variation of this model based on \cite{wong1,rl}.
 We assume that particles are produced in each inelastic
nucleon-nucleon collision. When traversing the nucleus, a nucleon
loses its energy. We assume a simple law according to which
the momentum of nucleon in the CMS nucleon-nucleon system is reduced by 
by a factor 1.7 ($p_{old}/p_{new}=1.7$)
in each collision. This corresponds to the lost of
about 0.5 units of  rapidity per collision.
The number of charged particles (normalized to 1. at $\sqrt(s)=20.$)
emitted in each
collision is then calculated according to the empirical rule:
$$
n(\sqrt(s))=\ln(\sqrt(s)/2.3)/\ln(8.7)
$$
where $n(2.3)=0.$ and $n(20)=1.$ , note that $\sqrt(s)=20 GeV$ is the SPS
heavy ion energy for sulphur-sulphur collision  and $\sqrt(s)=17 GeV$ is SPS energy
for lead-lead collision.

In fig. 3 the differential cross section of the quantity
\be 
w=\sum_{collisions} n(s)
\ee
is plotted. The choice of this quantity is motivated
by the fact that if there is no degradation of the energy
the quantity $w$ is equal to the number of collisions.
Comparison of fig.3 and fig.2 shows that the modified model
is also able to reproduce the data.

\begin{figure}[t]
\begin{center}\mbox{\input epsf \epsfysize 7cm
                        \epsfbox{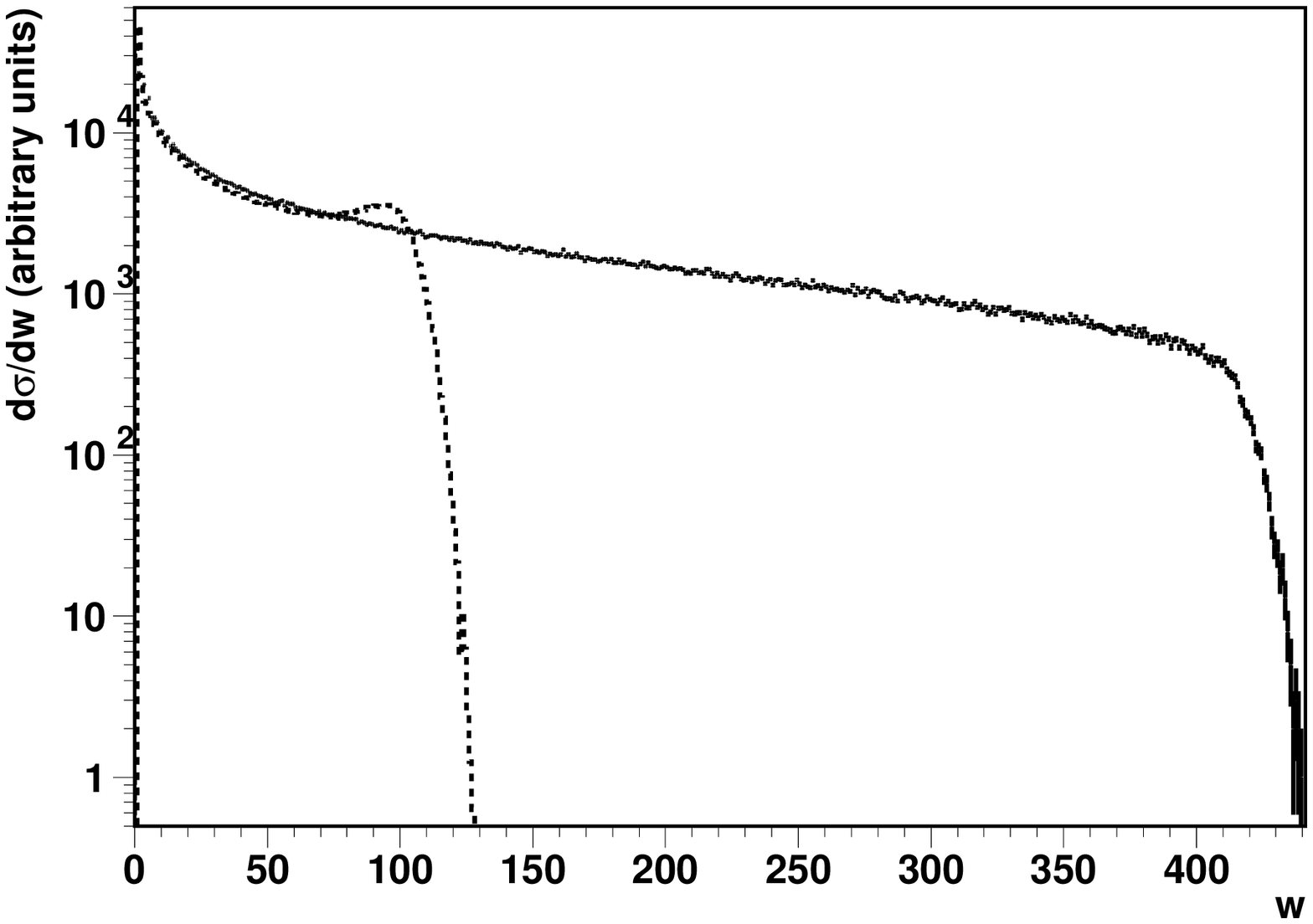}}\end{center}
\caption{Fig1 3. Differential cross section of the transverse energy
in modified model, for explanation see text.}
\end{figure}

\vspace{10pt}
\section{Conclusions}
\vspace{6pt}

We show that the transverse energy and charged particle
distributions of heavy ion collisions are determined
by the geometry of nuclei collisions. Assuming that the number
of produced particles and the transverse energy are
proportional to the number of participants the model is able
to describe the data for different heavy ion systems.

This model allows us to estimate the number of participants and 
impact parameters corresponding to different 
centrality triggers \cite{rlovb}.
Models based on the same principle were used to quantify
the centrality of the collision in the WA97 \cite{wa97} and
NA50 \cite{kluberg} collaborations. 
(The NA49 collaboration uses a different approach as they are able to 
measure the number of participants using
either a zero degree calorimeter 
 or the number of measured protons \cite{gazd}.

The phenomenological observation that the number of produced
particles is proportional to the number of participants
can be obtained in a modified model where 
\begin{enumerate}
\item nucleons loose their energy in each collision when traversing
the nucleus
\item particles are produced in each nucleon-nucleon collision
in amount dependent on the CMS energy of the nucleon-nucleon
collision.
\end{enumerate}
The extension of the modified model to include
the rapidity and transverse momentum spectra of produced
particles is straightforward. 

\vspace{10pt}
\section{Acknowledgement}
\vspace{6pt}
I am indebted to Orlando Villalobos Baillie for many useful
comments to the manuscript of this paper.


\begin{thebibliography}{99}

\bibitem{bialas}\refer{A. Bia\l as, A. Bleszynski, M.Czyz }{Nucl.Phys.B}{111}{1976}{461}.
\bibitem{ftak}\refer{N. Pi\v s\'utov\'a,P.Lichard, J.Pi\v s\'ut}
{Phys.Lett.B}{172}{1986}{451}\\
\refer{N. Pi\v s\'utov\'a, J.Pi\v s\'ut}
{Acta Physica Polonica B}{18}{1987}{177},\\
\refer{J.Ft\'a\v cnik, K.Kajantie, N.Pi\v s\'utov\'a, J. Pi\v s\'ut}
{Phys. Lett. B}{196}{1987}{387}
\bibitem{baym}\refer{G.Baym, P.Braun-Munziger,V.Ruuskanen}{Phys.Lett.B}
{190}{1987}{29}
\bibitem{jb}\refer{A.D.Jackson,H.Bogild}{Nucl.Phys.A}{470}{1987}{669}
\bibitem{khw}\refer{M.Kutschera, J.H\"ufner, K.Werener}{Physs.Lett. B}{192}{1987}{237}
\bibitem{sh}\refer{R.Shanta,S.K.Gupta}{Z.Phys.A}{338}{1991}{183}
\bibitem{wong1}\refer{J.Y.Zhang et al.}{Phys. Rev. C}{46}{1992}{748}
\bibitem{satz}\refer{D. Kharzeev et al.}{Z.Phys.C}{74}{1997}{307}
\bibitem{jean}\refer{S.Jean,J.Kapusta}{Phys. Rev. C}{56}{1997}{468}
\bibitem{rl}\refer{R.Lietava, J. Pi\v s\'ut}{Eur.Phys.J C}{5}{1998}{135}
\bibitem{the} R.J.Glauber, in {\sl Lecures in Theoretical Physics}, edited by
W.E.Brittin et al., Intersciemce, New York,1959, Vol. I.,p.315,\\
\refer{K.C.Chung,C.S.Wang,A.J.Santiago, G.Pech}{Phys. Rev. C}{57}{1998}{847}
\bibitem{wong}C.Y.Wong: {\sl Introduction to High-Energy Heavy Ion Collisions}, World
Scientific,1994
\bibitem{bachler}\refer{J. B\"achler et al.,}{Z. Phys. C}{51}{1991}{157}
\bibitem{DPM}\refer{A. Capella and J. Tran Thanh Van,}{Phys.Lett.B}{93}{1980}{146}
\bibitem{ivan} I. Kr{\'a}lik, Ph.D. Thesis, Institute of 
Experimental Physics, SAS, Ko{\v s}ice, 1995.
\bibitem{blasiuk}B. Lasiuk at al. {\sl Hadronic observanles at NA49}
School and Workshop on Heavy
Ion Collisions, September 1.-5., 1996, Bratislava, Slovakia. NA49 note 117;
\bibitem{na49}\refer{S.Margetis et al.}{Phys.Rev.Lett.}{75}{1995}{3814} 
\bibitem{rlovb}R.Lietava and O. Villalobos Baillie, {\sl A simple model
for average multiplicity}, WA97 internal note, July,1996;
\bibitem{wa97}\refer{E.Andersen et al.}{Phys. Lett. B}{433}{1998}{209}
\bibitem{kluberg}\refer{C. Baglin et al.}{Phys. Lett. B} {251}{1990}{472}
\bibitem{gazd}\refer{J. B\"achler et al.}{Phys.Rev.Lett.} {72}{1994}{1419}
\end{thebibliography}
\end{document}